\documentclass[sigconf]{acmart} 
\AtBeginDocument{%
  }

\setcopyright{acmlicensed}
\copyrightyear{2025}
\acmYear{2025}
\acmDOI{XXXXXXX.XXXXXXX}
\acmConference[GenAIRecP@KDD 2025]{In Proceedings of Second Workshop on Generative AI for Recommender Systems and Personalization}{August 4th,
  2025}{Toronto, Canada}
\acmISBN{978-1-4503-XXXX-X/2018/06}
\usepackage{pdflscape}
\usepackage{multirow}
\usepackage{adjustbox}
\usepackage{tcolorbox}
\usepackage{verbatim}
\usepackage{enumitem}
\usepackage{float}
\usepackage{graphicx}
\usepackage{makecell}

\DeclareUnicodeCharacter{25AA}{\textbullet}
\begin{document}

\title{LLM-Enhanced Reranking for Complementary Product Recommendation}




\author{Zekun Xu}
\authornote{Both authors contributed equally to this research.}
\affiliation{%
 \institution{North Carolina State University}
 \city{Raleigh}
 \state{North Carolina}
 \country{USA}
}

\author{Yudi Zhang}
\authornotemark[1]
\affiliation{%
 \institution{Iowa State University}
 \city{Ames}
 \state{Iowa}
 \country{USA}
}


\begin{abstract}
 Complementary product recommendation, which aims to suggest items that are used together to enhance customer value, is a crucial yet challenging task in e-commerce. While existing graph neural network (GNN) approaches have made significant progress in capturing complex product relationships, they often struggle with the accuracy-diversity tradeoff, particularly for long-tail items. This paper introduces a model-agnostic approach that leverages Large Language Models (LLMs) to enhance the reranking of complementary product recommendations. Unlike previous works that use LLMs primarily for data preprocessing and graph augmentation, our method applies LLM-based prompting strategies directly to rerank candidate items retrieved from existing recommendation models, eliminating the need for model retraining. Through extensive experiments on public datasets, we demonstrate that our approach effectively balances accuracy and diversity in complementary product recommendations, with at least 50\% lift in accuracy metrics and 2\% lift in diversity metrics on average for the top recommended items across datasets.
\end{abstract}




\maketitle

\section{Introduction}



In the era of e-commerce, effective product recommendation systems have become crucial for both customer satisfaction and business success. 
While traditional recommendation systems excel at suggesting similar or substitute products, recommending complementary products - items that are used together to create enhanced value for customers (e.g., a camera and its lens, or a printer and ink cartridges) — poses greater challenges and opportunities. Such recommendations require a deeper understanding of product relationships, use cases, and customer intent.
Accurately identifying and recommending complementary products is crucial for improving user experience and maximizing customer lifetime value. Yet traditional models often fall short, especially for long-tail products or in cold-start scenarios where historical purchase data is limited.

In recent years, graph neural network (GNN) models have demonstrated significant promise in addressing the challenges of complementary product recommendation \citep{hamilton2017inductive,xu2018powerful,liu2020,zhou2022,luo2025}. 
These models effectively capture the complex relationships between products by representing them as nodes in a heterogeneous graph structure, where edges represent various types of interactions and complementary relationships. 
GNNs can learn rich node embeddings by aggregating information from neighboring nodes through multiple message-passing layers, enabling them to capture both structural and semantic complementarity patterns. Recent approaches have incorporated attention mechanisms to weigh the importance of different neighbor relationships, while also leveraging side information such as product attributes, user behavior, and temporal dynamics. 
This has led to more accurate and contextually relevant complementary product recommendations, as the models can better understand both explicit and implicit relationships between items in the product network.
However, the challenge is still prominent in terms of balancing between popular and long-tail item recommendations, where GNNs still tend to recommend highly-connected products that leads to reduced novelty and diversity in complementary recommendations.

Latest advances in Large Language Models (LLMs) show potential to further improve complementary product recommendation by offering promising solutions to the accuracy-diversity tradeoff. 
LLMs, trained on vast amounts of text data, possess rich semantic understanding and can comprehend complex product relationships beyond simple co-occurrence patterns. 
Their ability to process and understand natural language descriptions, technical specifications, and user-generated content enables more sophisticated complementary product discovery and ranking mechanisms.
Through their pre-trained knowledge of product descriptions, use cases, and contextual relationships, LLMs can identify novel and meaningful complementary relationships that might not be apparent in traditional usage patterns or graph structures. 
This capability allows them to recommend less popular but highly relevant complementary items by understanding deeper semantic connections and functional relationships between products. 

There is recent literature on applying LLMs to enhance complementary product recommendation, but they primarily approach the problem through using LLMs to augment incomplete features or relationships in the product graph \citep{lyu2023llm,wei2024,wang2025}. 
While these approaches leverage the knowledge base from LLMs for enriching the input data, a drawback is that they would involve retraining the corresponding downstream graph recommendation models according to the revised input.
Moreover, there is no guarantee that the useful information from LLM-augmented input data will effectively flow to the model output depending on the architecture of the model.
Instead of using LLMs as a text or data preprocessor as in previous works, we propose to directly use LLMs to enhance the reranking on the retrieved items from recommendation models.
In the proposed model-agnostic approach, any graph recommendation model can be used without additional retraining overhead as a baseline filter to retrieve an initial candidate list of complementary products, on top of which LLMs are utilized to further improve the reranking based on accuracy and diversity criteria.
This idea of LLM-based reranking has already been explored in other domains \citep{hou2024,carraro2024,gao2025}, but to the best of our knowledge this work is the first attempt to adapt LLM prompting strategies to enhance reranking in complementary product recommendation by tackling the accuracy-diversity tradeoff.
Our key contributions include (code \footnote{\url{https://anonymous.4open.science/r/llm_rerank-4B01/README.md}}):
\begin{itemize}
    \item We conduct the first study to leverage LLM-based prompting strategy with multiple agents to enhance reranking algorithms for complementary product recommendation.
    \item We demonstrate the effectiveness of the proposed method through extensive experiments on four public datasets and provide in-depth insights into the performance metrics.
\end{itemize}

\section{Related Work}

\citet{mcauley2015} is one of the earliest work that formulates complementary product recommendation as a link prediction task, which predicts the relationships between pairs of products from associated description and review texts.
Under the same formulation, a variety of deep learning methods have since then been developed and applied to complementary product recommendation, including ENCORE \cite{Zhang2018}, Linked Variational Autoencoder \cite{Rakesh2019}, P-Companion \cite{Hao2020}, Decoupled Graph Convolution Network \cite{liu2020}, Graph Attention Network \cite{yan2022}, DAEMON \cite{virinchi2022recommending}, Decoupled Hyperbolic Graph Attention Network \cite{zhou2022}, Dynamic Policy Network \cite{yang2022}, Generative Adversarial Networks \cite{bibas2023}, and Spectral-based graph neural networks \cite{luo2025}.

With the advent of generative AI, there have been works that apply LLMs to complementary product recommendation, but most of them use LLMs to augment the input data through rewriting incomplete text description \cite{lyu2023llm}, enriching features in the graph \cite{wei2024}, or providing additional side information \citep{huang2023,wang2025}. 
Although LLMs can enrich input data, this approach has limitations as it requires model retraining and doesn't guarantee effective information transfer from the augmented input to the final output due to model architecture constraints
Outside the domain of complementary product recommendation, there is literature on using LLMs to directly enhance reranking method via prompt engineering \citep{hou2024,carraro2024,gao2025}.

\section{Problem Statement}

We formulate complementary product recommendation as a link prediction task in a product graph, which is commonly adopted in the literature \citep{hamilton2017inductive,xu2018powerful,liu2020,zhou2022,luo2025}.
Let $\mathcal{G}=\{\mathcal{V}, \mathcal{X}, \mathcal{E}\}$ denote the complementary product graph, where: $\mathcal{V}=\{v_1,\ldots,v_n\}$ is the set of items (nodes); $\mathcal{X}=\{\mathbf{x_1},\ldots,\mathbf{x_n}\}$ is the set of $d$-dimensional feature vectors for each of the nodes; $\mathcal{E}=\{e_{ij}\}$ is the set of undirected edges (links) between nodes, which represents complementary relationships. 
For a complementary product graph $\mathcal{G}$, we want to predict the probability of an edge $e_{ij}$ when given two items $v_i$ and $v_j$ so as to determine whether they are complementary.

 \begin{figure*}[!ht]
    \centering
    \includegraphics[width=0.95\textwidth]{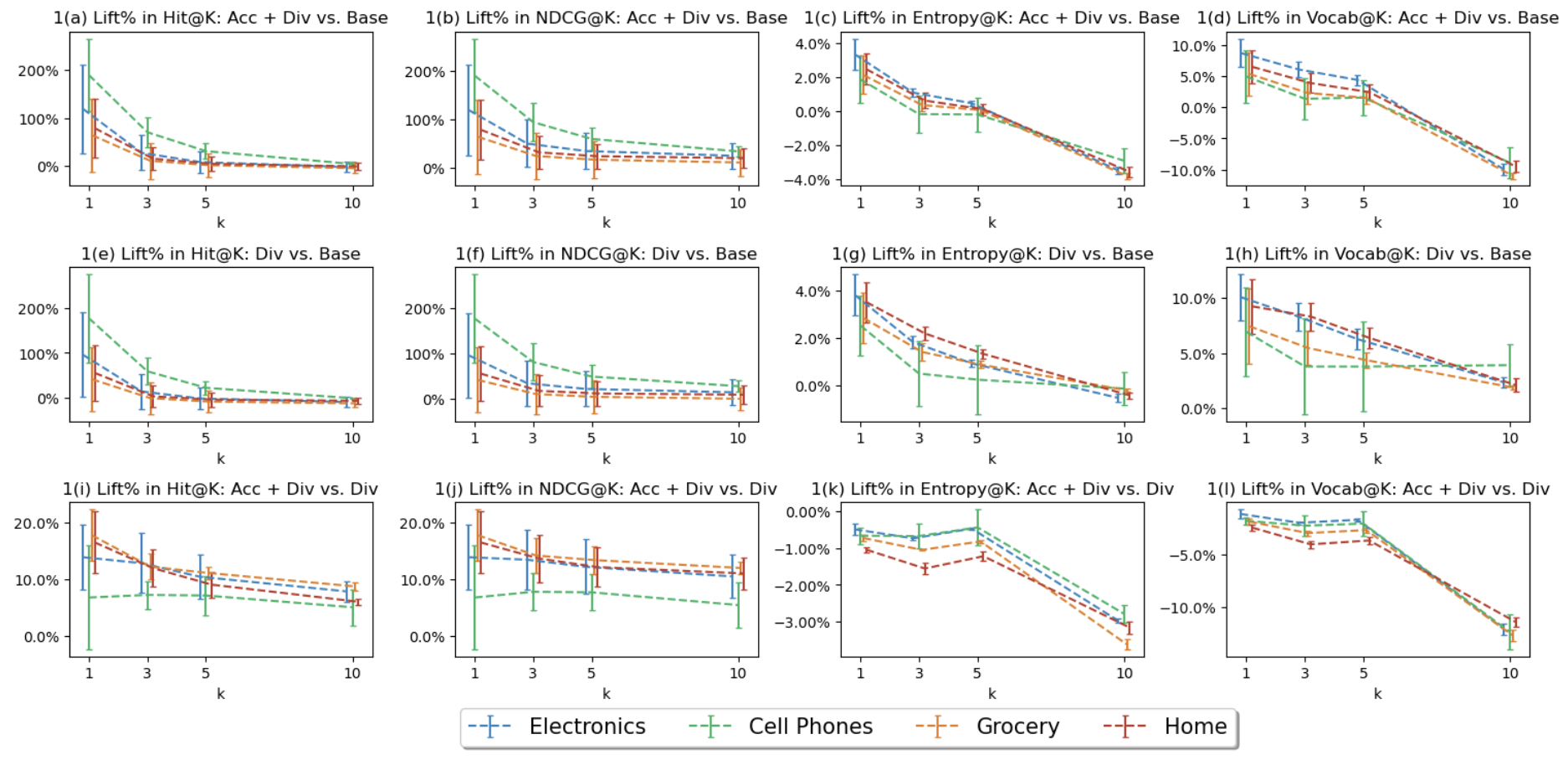}
    \caption{Lift in accuracy (Column 1: Hit; Column 2: NDCG) and diversity (Column 3: Entropy; Column 4: Vocabulary Size) metrics by dataset (Electronics, Cell Phones, Grocery, Home), where the standard error bar represents variability across three baseline GNNs (GraphSage, GAT, SComGNN). Row 1: overall enhancement with both diversity and accuracy agents vs. baseline; Row 2: ablation enhancement with diversity agent vs. baseline; Row 3: ablation enhancement with both diversity and accuracy agents vs. diversity agent only. The underlying LLM is Llama3.3-70B. Hyperparameter is 50 for diversity agent and 25 for accuracy agent.} 
    \label{result_plot_gnn}
\end{figure*}

\section{Method}

\subsection{Baseline Recommendation Model}

Without loss of generality, let $f_\mathcal{G}: \mathcal{X}\times\mathcal{X}\to\mathbb{R}$, denote a baseline recommendation model trained on the complementary product graph $\mathcal{G}$, which can take input from any pair of product feature vectors and produce a relevance score representing how likely there exists a complementary relationship between the two products.
One of the key advantages of our LLM-enhanced reranking approach versus previous works on LLM-based input data enrichment \citep{lyu2023llm,wei2024,wang2025} is that our approach is agnostic of the baseline recommendation model and does not invoke any model architecture change or retraining.
The role of the baseline recommendation model in our framework is a retriever, which performs an initial filtering on the large pool of potential candidates to top K (e.g. 50, but still much smaller than the size of the original pool).
Any model $f_{\mathcal{G}}$ that can produce relevance scores between complementary product pairs is eligible as the retriever, but the amount of improvement from later LLM reranking steps may differ depending on the quality of the baseline retriever.

\subsection{Enhanced Reranking on Diversity}

We decompose the proposed LLM-enhanced reranking into to two subflows: 1/ a `diversity agent' whose task is to improve reranking relevance from diversity perspective, and 2/ an `accuracy agent' whose task is to further improve reranking accuracy on top of the diversified list.
The diversity agent uses the retrieved item list from baseline recommendation model as input to refine the reranking, where the prompting strategy is structured as follows.

\paragraph{Input Format.}  

\noindent Considering a product, its basic information is:  
\begin{verbatim}
{title: XXXXXXXX}
\end{verbatim}

\vspace{0.5em}
\noindent Here's a list of the candidate products:  
\begin{itemize}[leftmargin=2em]
  \item[ID:0] title: XXXXXXXX
  \item[ID:1] title: XXXXXXXX
   \item[...]
\end{itemize}

\paragraph{Task Definition.}  
\noindent The task is identifying the complementary relation between the given product and candidates.  
Complementary is defined as: products are likely to be purchased or used at the same time, but it is not a direct substitute.

\paragraph{Few-shot Learning Examples.}  
\noindent A complementary product can be:
\begin{itemize}[leftmargin=2em]
  \item An accessory of the given product (e.g., iPhone Case is complementary to iPhone)
  \item Both accessories to the same product (e.g., Speaker Cables can be complementary to Speaker Stands)
  \item Products used together for the same activity (e.g., Bowl can be complementary to Plate)
\end{itemize}

\paragraph{Ranking Instructions.}  
\noindent Then rerank the candidates based on above given information. The order of reranking result should represent how likely the candidate is a complementary product.

\noindent Meanwhile, focus on the diversity aspect (more items with different `genre' feature at the top of the list).

\paragraph{Output Format.}  
\noindent Your answer should \textbf{ONLY} rank all mentioned candidates ID, do \textbf{NOT} repeat or include Name. And omit anything else such as your thinking and decision-making process.

\noindent Example answer format for 5 candidates: \verb|[1, 4, 3, 0, 2]|

\subsection{Enhanced Reranking on Accuracy}

To further enhance the reranking of the recommended complementary products, the accuracy agent will use the refined subset of items generated by the diversity agent as input to focus on the improvement on precision. 
Specifically, the accuracy agent follows the same prompting structure as the diversity agent, except that its \textit{Ranking Instructions} is updated as
"Meanwhile, focus on the \textbf{accuracy} aspect (choose items that are most precisely and correctly complementary to the given product)."










\begin{table*}
\caption{Accuracy (Hit$@$K and NDCG$@$K) and diversity (Entropy$@$K and Vocabulary$@$K) metrics by dataset (Cell Phones, Electronics, Grocery, Home) and method. The nine methods include combinations of baseline GNN (GraphSage, GAT, SComGNN) and enhanced reranking (none, diversity, accuracy + diversity). The underlying LLM is Llama3.3-70B. }
\centering
\setlength{\tabcolsep}{3pt}
\tiny
\begin{adjustbox}{width=1\textwidth,center}
\begin{tabular}{|c|c|cccc|cccc|cccc|cccc|}
\hline
\multirow{3}{*}{Method} & \multirow{3}{*}{K} & \multicolumn{16}{c|}{Datasets} \\
\cline{3-18}
& & \multicolumn{4}{c|}{Cell Phones} & \multicolumn{4}{c|}{Electronics} & \multicolumn{4}{c|}{Grocery} & \multicolumn{4}{c|}{Home} \\
& & Hit & NDCG & Entropy & Vocab & Hit & NDCG & Entropy & Vocab & Hit & NDCG & Entropy & Vocab & Hit & NDCG & Entropy & Vocab \\
\hline
\multirow{4}{*}{\makecell{GraphSage \\ (Base)}} & 1 & .154 & .154 & 2.86 & 19.5 & .323 &  .323 & 2.59 & 15.2 & .320 & .320 & 2.29 & 10.9 & .383 & .383 & 2.35 & 11.7 \\
& 3 & .397 & .302 & 3.70 & 47.7 & .588 & .477 & 3.66 & 42.7 & .576 & .469 & 3.33 & 30.3 & .642 & .535 & 3.43 & 33.4 \\
& 5 & .580 & .384 & 4.03 & 70.9 & .702 & .525 &  4.13 & 68.7 & .693 & .518 & 3.79 & 48.4 & .743 &  .576 & 3.92 &  54.3\\
& 10 & .791 & .458 & 4.46 & 122.9 & .841 & .570 & 4.74 & 130.2 & .830 & .563 & 4.39 & 91.1 & .845 & .610 & 4.58 & 105.0 \\
\hline
\multirow{4}{*}{\makecell{GraphSage \\ (Div.)}} & 1 & .306 & .306 & 2.95 & 21.2 & .412 & .412 & 2.66 & 16.4 & .335 & .335 & 2.32 & 11.3 & .466 & .466 & 2.41 & 12.6 \\
& 3 & .530 & .444 & 3.73 & 49.7 & .534 & .483 & 3.71 & 45.6 & .485 & .422 & 3.37 & 31.4 & .601 & .544 & 3.50 & 35.7 \\
& 5 & .639 & .493 & 4.06 & 74.4 & .609 & .514 & 4.16 & 72.3 & .579 & .460 & 3.82 & 50.1 & .679 & .577 & 3.97 & 57.3 \\
& 10 & .779 & .543 & 4.44 & 127.5 & .730 & .552 & 4.71 & 132.8 & .713 & .503 & 4.39 & 93.0 & .783 & .610 & 4.55 & 106.8 \\
\hline
\multirow{4}{*}{\makecell{GraphSage \\ (Div.+Acc.)}} & 1 & .351 & .351 & 2.93 & 20.8 & .487 & .487 & 2.65 & 16.2 & .400 & .400 & 2.31 & 11.1 & .551 & .511 & 2.38 & 12.2 \\
& 3 & .580 & .495 & 3.71 & 48.7 & .622 & .566 & 3.69 & 44.8 & .546 & .485 & 3.33 & 30.4 & .680 & .626 & 3.44 & 34.2 \\
& 5 &  .704 & .549 & 4.04 & 72.9 & .690 & .594 & 4.14 & 71.2 & .645 & .525 & 3.79 & 48.6 & .751 & .655 & 3.92 & 55.1 \\
& 10 & .837 & .598 &  4.33 & 109.3 & .799 & .626 & 4.57 & 116.1 & .770 & .566 & 4.22 & 80.6 & .834 & .685 & 4.41 & 94.6 \\
\hline
\multirow{4}{*}{\makecell{GAT \\ (Base)}} & 1 & .126 & .126 & 2.85 & 19.4 & .271 & .271 & 2.57 & 15.0 & .331 & .331 & 2.25 & 10.4 & .390 & .390 & 2.33 & 11.5  \\
& 3 & .354 & .264 &  3.69 & 47.1 & .590 & .456 & 3.64 & 42.1 & .625 & .503 & 3.31 & 29.4 & .704 & .574 & 3.42 & 32.8  \\
& 5 & .550 & .350 & 4.04 & 70.7 & .753 & .523 & 4.12 & 68.0 & .762 & .559 & 3.77 & 47.3 & .821 & .622 & 3.91 & 53.4  \\
& 10 & .825 & .448 & 4.48 & 122.9 & .909 & .574 & 4.74 & 129.9 & .901 & .605 & 4.37 & 89.4 & .918 & .654 & 4.56 & 103.6  \\
\hline
\multirow{4}{*}{\makecell{GAT \\ (Div.)}} & 1 & .309 & .309 & 2.95 & 21.2 & .425 & .425 & 2.67 & 16.5 & .318 & .318 & 2.33 & 11.3 & .463 & .463 & 2.41 & 12.6  \\
& 3 & .539 & .446 & 3.75 & 50.8 & .548 & .497 & 3.71 & 45.7 & .460 & .400 & 3.36 & 30.2 &  .597 & .540 & 3.49 & 35.4  \\
& 5 & .649 & .494 & 4.09 & 75.8 & .620 & .526 & 4.16 & 72.3 & .559 & .441 & 3.81 & 49.5 &  .684 & .576 & 3.96 & 56.7  \\
& 10 & .811 & .549 & 4.51 & 130.1 & .746 & .566 & 4.71 & 132.7 & .716 & .491 & 4.37 & 91.1 & .800 & .613 & 4.54 & 105.5  \\
\hline
\multirow{4}{*}{\makecell{GAT \\ (Div.+Acc.)}} & 1 & .337 & .337 & 2.93 & 20.9 & .494 & .494 & 2.65 & 16.3 & .386 & .386 & 2.31 & 11.1 & .560 & .560 & 2.38 & 12.2  \\
& 3 & .564 & .477 & 3.71 & 49.1 & .632 & .576 & 3.69 & 44.7 & .526 & .468 & 3.32 & 30.2 & .684 & .632 & 3.43 & 34.0  \\
& 5 & .671 & .526 & 4.05 & 73.3 & .695 & .602 & 4.14 & 70.9 & .627 & .509 & 3.78 & 48.1 & .754 & .661 & 3.91 & 54.5  \\
& 10 & .823 & .567 & 4.38 & 115.3 & .809 & .634 & 4.57 & 116.7 & .779 & .554 & 4.21 & 80.0 & .850 & .693 & 4.39 & 93.0  \\
\hline
\multirow{4}{*}{\makecell{SComGNN \\ (Base)}} & 1 & .087 & .087 & 2.90 & 20.3 & .152 & .152 & 2.53 & 14.4 & .179 & .179 & 2.24 & 10.2 & .232 & .232 & 2.30 & 11.1 \\
& 3 & .274 & .198 & 3.76 & 49.8 & .376 & .281 & 3.63 & 41.2 & .397 & .305 & 3.30 & 29.0 & .495 & .384 & 3.40 & 32.0 \\
& 5 & .458 & .281 & 4.11 & 74.2 & .519 & .340 & 4.11 & 66.9 & .530 & .360 & 3.76 & 46.7 & .627 & .439 & 3.89 & 52.3 \\
& 10 & .753 & .383 & 4.52 & 125.6 & .707 & .401 & 4.74 & 128.3 & .727 & .424 & 4.37 & 88.7 & .772 & .486 & 4.54 & 101.4 \\
\hline
\multirow{4}{*}{\makecell{SComGNN \\ (Div.)}} & 1 & .337 & .337 & 2.93 & 20.8 & .460 & .460 & 2.65 & 16.2 & .401 & .401 & 2.32 & 11.3 & .525 & .525 & 2.40 & 12.4 \\
& 3 & .524 & .450 & 3.73 & 49.5 & .598 & .540 & 3.70 & 45.1 & .556 & .492 & 3.35 & 30.9 & .652 & .600 & 3.48 & 35.1 \\
& 5 & .636 & .498 & 4.05 & 73.6 & .657 & .566 & 4.15 & 71.6 & .627 & .521 & 3.80 & 49.0 & .712 & .624 & 3.95 & 56.2 \\
& 10 & .761 & .545 & 4.49 & 128.1 & .734 & .591 & 4.72 & 132.0 & .710 & .548 & 4.35 & 90.2 & .778 & .646 & 4.53 & 104.2 \\
\hline
\multirow{4}{*}{\makecell{SComGNN \\ (Div.+Acc.)}} & 1 & .326 & .326 & 2.91 & 20.4 & .494 & .494 & 2.64 & 16.0 & .452 & .452 & 2.31 & 11.1 & .580 & .580 & 2.38 & 12.2 \\
& 3 & .565 & .473 & 3.71 & 48.7 & .639 & .580 & 3.68 & 44.2 & .611 & .545 & 3.31 & 30.1 & .706 & .654 & 3.43 & 33.8 \\
& 5 & .687 & .525 & 4.05 & 72.9 & .697 & .604 & 4.13 & 70.4 & .689 & .577 & 3.76 & 47.8 & .758 & .676 & 3.91 & 54.3 \\
& 10 & .810 & .562 & 4.35 & 113.6 & .777 & .628 & 4.58 & 116.8 & .778 & .607 & 4.19 & 78.9 & .821 & .697 & 4.39 & 92.8\\
\hline
\end{tabular}
\end{adjustbox}
\label{table:all_results}
\end{table*} 

\section{Results}

\begin{figure*}[!ht]
    \centering
    \includegraphics[width=0.95\textwidth]{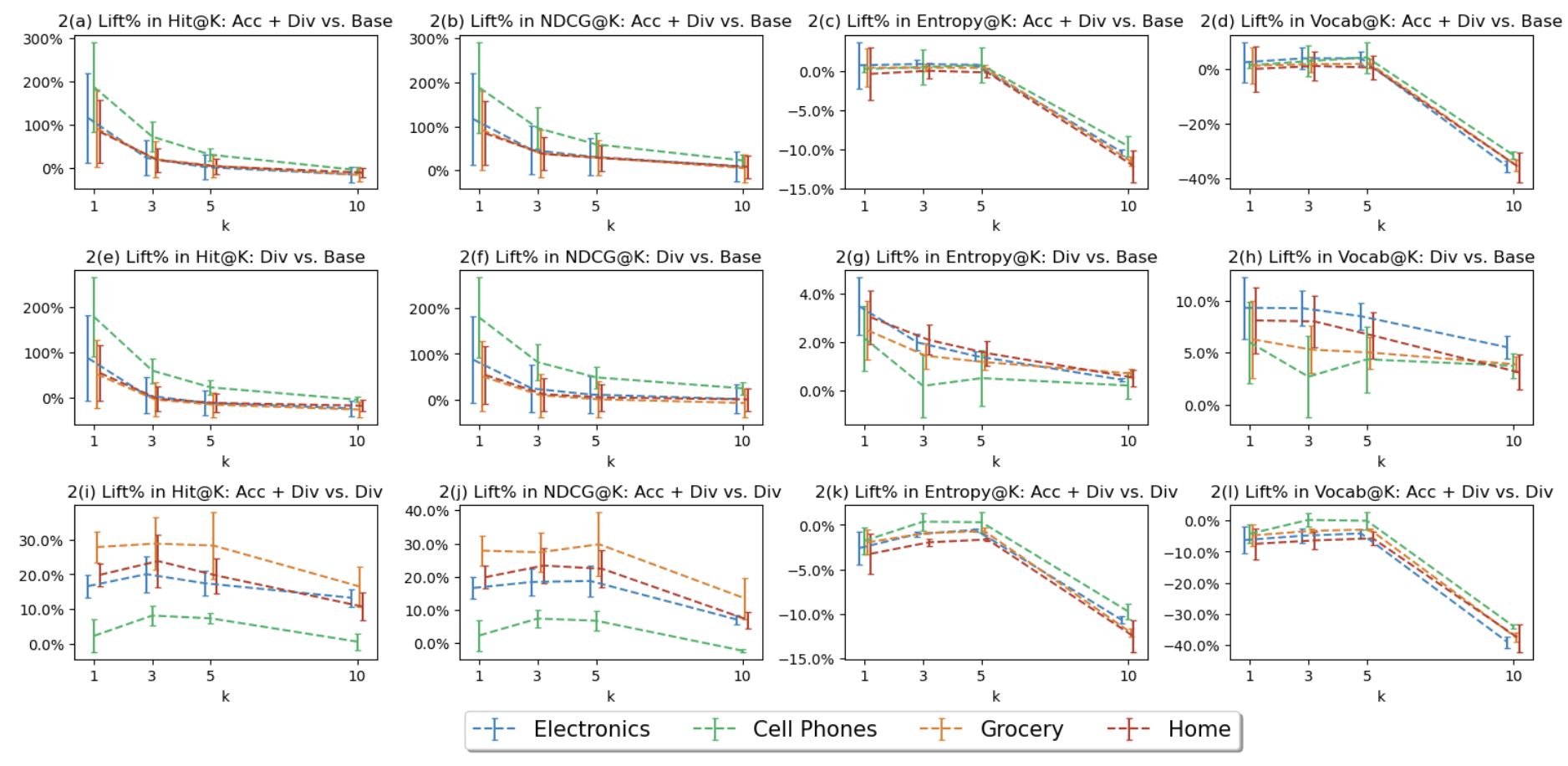}
    \caption{Lift in accuracy (Column 1: Hit; Column 2: NDCG) and diversity (Column 3: Entropy; Column 4: Vocabulary Size) metrics by dataset (Electronics, Cell Phones, Grocery, Home), where the standard error bar represents variability across three baseline GNNs (GraphSage, GAT, SComGNN). Row 1: overall enhancement with both diversity and accuracy agents vs. baseline; Row 2: ablation enhancement with diversity agent vs. baseline; Row 3: ablation enhancement with both diversity and accuracy agents vs. diversity agent only. The underlying LLM is Llama3.3-70B. Hyperparameter is 100 for diversity agent and 50 for accuracy agent.} 
    \label{result_plot_gnn_100}
\end{figure*}

We use Amazon product review data\footnote{\url{https://jmcauley.ucsd.edu/data/amazon/}} across four consumer categories: \textit{Electronics}, \textit{Cell Phones}, \textit{Grocery}, and \textit{Home} for experiments. 
We consider three graph neural network models (GNNs) as the baseline retriever: GraphSAGE~\cite{hamilton2017}, 
Graph Attention Network (GAT)~\cite{velivckovic2017},
and Spectral-based Complementary Graph Neural Networks (ScomGNN)~\cite{luo2025},
which serves to extracting an initial list of top 50 complementary product candidates. 
All three baseline GNN models are trained on the same graph datasets, where node features only include multi-level product categories and pricing.
During the diversity enhancement stage, the top 50 products ranked by each GNN retriever along with their unstructured product title texts are passed to a diversity agent powered by Llama3.3-70B~\cite{touvron2024llama3} model, where the agent is encouraged to select  across a broader range of product types.
Finally, the top 25 products obtained from the diversity-enhanced list are further refined by a Llama3.3-70B-powered accuracy agent to focus on recommendation precision.
Note that 50 and 25 are hyperparameters for the diversity and accuracy agents in Figure \ref{result_plot_gnn} and Table \ref{table:all_results}. 
In Figure \ref{result_plot_gnn_100} and supplementary material, we evaluate a different configuration of hyperparameters with 100 for diversity and 50 for accuracy agent, which shows similar results.

We compute standard accuracy metrics including: 1/ hit rate, and 2/ normalized discounted cumulative gain (NDCG), as well as exploratory diversity metrics including: 1/ size of vocabulary in product title, 2/ entropy of the vocabulary in product title distribution, which measures the diversity and randomness in the recommended items. A higher value indicates a more varied recommendation set. The entropy $H$ is calculated as:
$- \sum_{i=1}^{N} p_i \log p_i$, where $p_i$ is the probability of the $i$-th token appearing in the recommendation output, and $N$ is the total number of distinct tokens. 
All metrics use cutoff values $K \in \{1, 3, 5, 10\}$. 
Figure \ref{result_plot_gnn} shows the lift percentage in accuracy and diversity metrics by dataset and method, which summarizes the key insights from the full empirical results in Table \ref{table:all_results}.

In Row 1 in Figure 1, we compare the overall lift in accuracy metrics from both diversity and accuracy agents against averaging across baseline GNNs in terms of Hit$@$K in Figure \ref{result_plot_gnn}(a) and NDCG$@$K in \ref{result_plot_gnn}(b), we see a mean lift close to 200\% at $K=1$ in Cell Phones, followed by about 100\% lift in Electronics, and about 50\% lift in Home and Grocery. 
The higher lift in Cell Phone is primarily due to its lower baseline GNN accuracy, and that the lift decreases as K increases with a smaller gap between datasets.
Figure \ref{result_plot_gnn}(c) and \ref{result_plot_gnn}(d) capture the lift in diversity metrics, which shows a mean lift of at least 2\% in entropy and 5\% in vocabulary size at $K=1$. 
As $K$ increases, the lift in diversity metrics can be negative, which indicates the accuracy-diversity tradeoff: while there is potential for LLM-enhanced reranking to improve both accuracy and diversity metrics at smaller values of $K$, further improvements in accuracy will come at the cost of loss in diversity metrics at larger values of $K$.
Next, we decompose the effect separately from diversity and accuracy agent through ablation studies to quantify how individual agents contribute to the overall lift.

\subsection{Ablation Results on Diversity Agent}

Row 2 in Figure \ref{result_plot_gnn} presents ablation results from diversity agent alone against baseline GNN model, which shows diversity agent can significantly improve both accuracy and diversity metrics at smaller values of $K$.
Specifically, lift percentages in accuracy metrics of Hit$@$K in Figure \ref{result_plot_gnn}(e) and NDCG$@$K in \ref{result_plot_gnn}(f) are at a similar scale to the overall lift in Figure \ref{result_plot_gnn}(a) and \ref{result_plot_gnn}(b), which is at least greater than 50\% on average.
For diversity metrics in terms of textual entropy in Figure \ref{result_plot_gnn}(g) and vocabulary size in Figure \ref{result_plot_gnn}(h), the mean lift is greater than 2\% in entropy and 5\% in vocabulary size at $K=1$, with the smallest lift in Cell Phone because its baseline entropy and vocabulary size are larger than the other datasets and that the gap decreases with increasing $K$.
Interestingly, although diversity metrics still decrease as $K$ increases, they do not become negative as comparing to Figure \ref{result_plot_gnn}(c) and \ref{result_plot_gnn}(d), which indicates that there is opportunity to simultaneously improve both accuracy and diversity metrics on top of baseline through diversity-enhanced reranking.

\subsection{Ablation Results on Accuracy Agent}

 Row 3 in Figure \ref{result_plot_gnn} presents ablation results from accuracy agent on top of enhanced ranking list produced by diversity agent, which shows that accuracy agent can still significantly improve accuracy but it comes at the cost of decreasing diversity metrics.
Specifically, Figure \ref{result_plot_gnn}(i) and \ref{result_plot_gnn}(j) compare the gain in Hit$@$K and NDCG$@$K from accuracy agent on top of diversity agent, which shows accuracy agent can further increase both hit rate and NDCG by at least 5\% on average across models in all datasets. 
This additional lift from accuracy agent is the smallest in Cell Phones because the previous reranking step from diversity agent already improves accuracy metrics in Cell Phones more than the other datasets. 
Figure \ref{result_plot_gnn}(k) and \ref{result_plot_gnn}(l) shows that accuracy agent will decrease both diversity metrics of entropy and vocabulary size from the diversity-enhanced reranking list.
The findings from Figure \ref{result_plot_gnn}(i) - \ref{result_plot_gnn}(l) are expected because of the accuracy-diversity tradeoff: at this stage there is no "free lunch" as the further increment in reranking accuracy will have to come at the cost of loss in diversity.

\subsection{Ablation Results on Agent Hyperparameter}

Figure \ref{result_plot_gnn_100} has similar construction and interpretation as Figure \ref{result_plot_gnn}, where the only difference is regarding the hyperparameter setting in diversity and accuracy agents.
The diversity agent now reranks the top 100 items, where the accuracy agent further refines the top 50 items rom the diversified item list.
The insights and results are similar as shown in Figure \ref{result_plot_gnn}.
Row 1 in Figure \ref{result_plot_gnn_100} is the overall effect expected by the accuracy-diversity tradeoff.
From Figure \ref{result_plot_gnn_100}(e) - \ref{result_plot_gnn_100}(h) in Row 2, we see that diversity agent is able to improve both accuracy metrics and diversity metrics versus baseline GNN model output.
From Figure \ref{result_plot_gnn_100}(i) - \ref{result_plot_gnn_100}(l) in Row 3, we see that accuracy agent is able to further improve accuracy metrics based on the reranked list produced by diversity agent, but this comes at the cost of decrease in diversity metrics.

\section{Conclusion}

We proposed an LLM-enhanced reranking method for complementary product recommendations, which has demonstrated promising results in addressing the accuracy-diversity tradeoff across multiple real-world datasets. 
In particular, our results indicate that the proposed diversity agent can enhance both accuracy and diversity ranking metrics on top of baseline GNN output, whereas further improvement in reranking accuracy induced by the proposed accuracy agent comes at the cost of decreasing diversity metrics.
The proposed method is a special case of a multi-agent system where there is only one interaction between accuracy and diversity agent in the reranking process, which we plan to extend in future work into an iterative multi-agent collaboration system where agents interact and learn from each other across multiple rounds, enabling continuous improvement in their performance over time.

\bibliography{ref}
\bibliographystyle{plainnat}

\onecolumn
\newpage

\vspace{1in}

\section*{Supplementary Material}

\begin{table*}[h]
\caption{Accuracy (Hit$@$K and NDCG$@$K) and diversity (Entropy$@$K and Vocabulary$@$K) metrics by dataset (Cell Phones, Electronics, Grocery, Home) and method. The nine methods include combinations of baseline GNN (GraphSage, GAT, SComGNN) and enhanced reranking (none, diversity, accuracy + diversity). The underlying LLM is Llama3.3-70B. The diversity agent reranks the top 100 items (as opposed to the top 50 items in the main paper), where the accuracy agent further refines the top 50 items (as opposed to the top 25 items in the main paper) from the diversified item list.}
\centering
\setlength{\tabcolsep}{3pt}
\tiny
\begin{adjustbox}{width=1\textwidth,center}
\begin{tabular}{|c|c|cccc|cccc|cccc|cccc|}
\hline
\multirow{3}{*}{Method} & \multirow{3}{*}{K} & \multicolumn{16}{c|}{Datasets} \\
\cline{3-18}
& & \multicolumn{4}{c|}{Cell Phones} & \multicolumn{4}{c|}{Electronics} & \multicolumn{4}{c|}{Grocery} & \multicolumn{4}{c|}{Home} \\
& & Hit & NDCG & Entropy & Vocab & Hit & NDCG & Entropy & Vocab & Hit & NDCG & Entropy & Vocab & Hit & NDCG & Entropy & Vocab \\
\hline
\multirow{4}{*}{\makecell{GraphSage \\ (Base)}} & 1 & .154 & .154 & 2.86 & 19.5 & .323 &  .323 & 2.59 & 15.2 & .320 & .320 & 2.29 & 10.9 & .383 & .383 & 2.35 & 11.7 \\
& 3 & .397 & .302 & 3.70 & 47.7 & .588 & .477 & 3.66 & 42.7 & .576 & .469 & 3.33 & 30.3 & .642 & .535 & 3.43 & 33.4 \\
& 5 & .580 & .384 & 4.03 & 70.9 & .702 & .525 &  4.13 & 68.7 & .693 & .518 & 3.79 & 48.4 & .743 &  .576 & 3.92 &  54.3\\
& 10 & .791 & .458 & 4.46 & 122.9 & .841 & .570 & 4.74 & 130.2 & .830 & .563 & 4.39 & 91.1 & .845 & .610 & 4.58 & 105.0 \\
\hline
\multirow{4}{*}{\makecell{GraphSage \\ (Div.)}} & 1 & .323 & .323 & 2.93 & 20.9 & .388 & .388 & 2.65 & 16.2 & .363 & .363 & 2.31 & 11.1 & .466 & .466 & 2.39 & 12.3 \\
& 3 & .543 & .459 & 3.72 & 49.3 & .476 & .440 & 3.72 & 45.9 & .455 & .417 & 3.36 & 31.1 & .557 & .520 & 3.48 & 35.2 \\
& 5 & .647 & .505 & 4.07 & 74.5 & .525 & .460 & 4.18 & 73.7 & .508 & .439 & 3.82 & 50.0 & .608 & .541 & 3.97 & 56.8 \\
& 10 & .753 & .543 & 4.49 & 128.7 & .575 & .475 & 4.76 & 136.5 & .561 & .456 & 4.42 & 94.0 & .661 & .559 & 4.58 & 106.7 \\
\hline
\multirow{4}{*}{\makecell{GraphSage \\ (Div.+Acc.)}} & 1 & .326 & .326 & 2.88 & 20.0 & .455 & .455 & 2.55 & 14.7 & .461 & .461 & 2.25 & 10.4 & .547 & .547 & 2.27 & 10.9 \\
& 3 & .578 & .488 & 3.76 & 50.4 & .587 & .532 & 3.67 & 43.0 & .601 & .530 & 3.33 & 29.9 & .687 & .651 & 3.41 & 32.2 \\
& 5 & .692 & .550 & 4.12 & 76.2 & .636 & .563 & 4.16 & 70.0 & .655 & .566 & 3.80 & 48.4 & .734 & .670 & 3.89 & 52.2 \\
& 10 & .714 & .564 & 4.08 & 85.2 & .641 & .586 & 4.24 & 80.4 & .667 & .499 & 3.88 & 57.4 & .705 & .577 & 3.91 & 60.7 \\
\hline
\multirow{4}{*}{\makecell{GAT \\ (Base)}} & 1 & .126 & .126 & 2.85 & 19.4 & .271 & .271 & 2.57 & 15.0 & .331 & .331 & 2.25 & 10.4 & .390 & .390 & 2.33 & 11.5  \\
& 3 & .354 & .264 &  3.69 & 47.1 & .590 & .456 & 3.64 & 42.1 & .625 & .503 & 3.31 & 29.4 & .704 & .574 & 3.42 & 32.8  \\
& 5 & .550 & .350 & 4.04 & 70.7 & .753 & .523 & 4.12 & 68.0 & .762 & .559 & 3.77 & 47.3 & .821 & .622 & 3.91 & 53.4  \\
& 10 & .825 & .448 & 4.48 & 122.9 & .909 & .574 & 4.74 & 129.9 & .901 & .605 & 4.37 & 89.4 & .918 & .654 & 4.56 & 103.6  \\
\hline
\multirow{4}{*}{\makecell{GAT \\ (Div.)}} & 1 & .315 & .315 & 2.94 & 21.1 & .396 & .396 & 2.66 & 16.3 & .337 & .337 & 2.32 & 11.2 & .448 & .448 & 2.40 & 12.4 \\
& 3 & .534 & .445 & 3.73 & 50.1 & .481 & .446 & 3.72 & 46.0 & .420 & .386 & 3.36 & 31.2 & .527 & .494 & 3.49 & 35.4 \\
& 5 & .631 & .487 & 4.09 & 75.8 & .526 & .465 & 4.18 & 73.8 & .472 & .407 & 3.82 & 50.0 & .581 & .516 & 3.97 & 56.8 \\
& 10 & .733 & .523 & 4.49 & 128.0 & .577 & .479 & 4.76 & 136.5 & .528 & .424 & 4.41 & 93.5 & .642 & .537 & 4.58 & 106.6 \\
\hline
\multirow{4}{*}{\makecell{GAT \\ (Div.+Acc.)}} & 1 & .310 & .310 & 2.85 & 19.6 & .473 & .473 & .256 & .15.0 & .447 & .447 & 2.25 & 10.5 & .554 & .554 & 2.30 & 11.3 \\
& 3 & .570 & .469 & 3.76 & 50.3 & .591 & .537 & 3.68 & 43.3 & .564 & .515 & 3.33 & 30.1 & .693 & .630 & 3.41 & 32.9 \\
& 5 & .688 & .503 & 4.12 & 76.3 & .617 & .559 & 4.16 & 70.3 & .637 & .569 & 3.80 & 48.7 & .721 & .656 & 3.90 & 53.8 \\
& 10 & .720 & .533 & 4.07 & 85.2 & .637 & .588 & 4.27 & 85.3 & .649 & .573 & 3.90 & 59.9 & .734 & .588 & 4.06 & 69.4 \\
\hline
\multirow{4}{*}{\makecell{SComGNN \\ (Base)}} & 1 & .087 & .087 & 2.90 & 20.3 & .152 & .152 & 2.53 & 14.4 & .179 & .179 & 2.24 & 10.2 & .232 & .232 & 2.30 & 11.1 \\
& 3 & .274 & .198 & 3.76 & 49.8 & .376 & .281 & 3.63 & 41.2 & .397 & .305 & 3.30 & 29.0 & .495 & .384 & 3.40 & 32.0 \\
& 5 & .458 & .281 & 4.11 & 74.2 & .519 & .340 & 4.11 & 66.9 & .530 & .360 & 3.76 & 46.7 & .627 & .439 & 3.89 & 52.3 \\
& 10 & .753 & .383 & 4.52 & 125.6 & .707 & .401 & 4.74 & 128.3 & .727 & .424 & 4.37 & 88.7 & .772 & .486 & 4.54 & 101.4 \\
\hline
\multirow{4}{*}{\makecell{SComGNN \\ (Div.)}} & 1 & .329 & .329 & 2.92 & 20.7 &  .449 & .449 & 2.65 & 16.2 & .427 & .427 & 2.32 & 11.2 & .525 & .525 & 2.39 & 12.4 \\
& 3 & .521 & .446 & 3.72 & 49.1 & .564 & .517 & 3.71 & 45.8 & .551 & .500 & 3.35 & 31.0 & .631 & .588 & 3.49 & 35.4 \\
& 5 & .638 & .494 & 4.07 & 74.8 & .612 & .537 & 4.17 & 73.4 & .609 & .524 & 3.81 & 49.6 & .684 & .609 & 3.97 & 57.0 \\
& 10 & .752 & .535 & 4.50 & 128.6 & .667 & .554 & 4.76 & 136.9 & .664 & .542 & 4.39 & 92.2 & .735 & .627 & 4.58 & 106.3 \\
\hline
\multirow{4}{*}{\makecell{SComGNN \\ (Div.+Acc.)}} & 1 & .354 & .354 & 2.91 & 20.5 & .508 & .508 & 2.63 & 16.0 & .529 & .529 & 2.31 & 11.1 & .622 & .622 & 2.38 & 12.1 \\
& 3 & .580 & .493 & 3.69 & 48.1 & .643 & .588 & 3.69 & 44.7 & .662 & .608 & 3.32 & 30.2 & .736 & .690 & 3.44 & 34.2 \\
& 5 & .677 & .534 & 4.04 & 72.5 & .697 & .610 & 4.14 & 71.4 & .721 & .632 & 3.77 & 48.2 & .781 & .708 & 3.91 & 54.9 \\
& 10 & .703 & .542 & 4.02 & 84.0 & .743 & .595 & 4.23 & 83.8 & .745 & .640 & 3.84 & 57.0 & .793 & .716 & 4.04 & 68.0\\
\hline
\end{tabular}
\end{adjustbox}
\label{table:all_results_100}
\end{table*}

\end{document}